\documentclass[prb,eqsecnum,twocolumn]{revtex4}

\usepackage{graphicx}%
\usepackage{dcolumn}
\usepackage{amsmath}
\usepackage{color}
\begin{document}

\newcommand{\be}{\begin{equation}}
\newcommand{\ee}{\end{equation}}
\newcommand{\nin}{\noindent}

\newcommand{\femn}{\ensuremath {\text{Fe}_{50}\text{Mn}_{50}\,}}
\newcommand{\nife}{\ensuremath{\text{Ni}_{81}\text{Fe}_{19}\,}}
\newcommand{\spini}{\ensuremath{\mathbf{S}_{i}}}

\newcommand{\hres}{\ensuremath{H_{res}\,}}
\newcommand{\kone}{\ensuremath{K^{(1)}_{i}\,}}
\newcommand{\ktwo}{\ensuremath{K^{(2)}_{i}\,}}
\newcommand{\field}{\ensuremath{\mathbf{H}\,}}
\newcommand{\alm}{\ensuremath{\alpha_{M}\,}}
\newcommand{\Na}{\ensuremath{N_{a}\,}}
\newcommand{\Nb}{\ensuremath{N_{b}\,}}
\newcommand{\nn}{\ensuremath{N_{b}/N\,}}
\newcommand{\nf}{\ensuremath{N_{F}\,}}
\newcommand{\naf}{\ensuremath{N_{AF}\,}}
\newcommand{\efaf}{\ensuremath{\epsilon_{F,AF}\,}}
\newcommand{\jaf}{\ensuremath{J_{AF}\,}}
\newcommand{\jfaf}{\ensuremath{J_{F,AF}\,}}
\newcommand{\hc}{\ensuremath{h_{c}\,}}
\newcommand{\hb}{\ensuremath{h_{b}\,}}
\newcommand{\ka}{\ensuremath{K_{a}\,}}
\newcommand{\kb}{\ensuremath{K_{b}\,}}
\newcommand{\keff}{\ensuremath{K^{eff}_{4}\,}}
\newcommand{\rtwo}{\ensuremath{R^{2}\,}}
\newcommand{\vh}{\ensuremath{\mathbf{h}\,}}
\newcommand{\vf}{\ensuremath{\mathbf{f}\,}}
\newcommand{\va}{\ensuremath{\mathbf{a}\,}}
\newcommand{\vb}{\ensuremath{\mathbf{b}\,}}
\newcommand{\marc}{\bf}
\newcommand{\hans}{\bf}
\preprint{Submitted to J. Appl. Phys.}

% Force line breaks with \\
\title[ ]{Separation of the first- and second-order contributions in magneto-optic Kerr effect magnetometry of epitaxial FeMn/NiFe bilayers}

\author{T. Mewes}
\email{mewes@mps.ohio-state.edu}
\affiliation{%
Department of Physics, 1077 Smith Laboratory\\
Ohio State University\\
174 W 18th Ave, Columbus, OH 43210, USA\\
}

\author{H. Nembach}
\affiliation{%
Fachbereich Physik and Forschungsschwerpunkt MINAS\\
Technische Universit{\"a}t Kaiserslautern\,\\
Erwin-Schr{\"o}dinger-Str.\ 56, 67663 Kaiserslautern, Germany\\
}

\author{M. Rickart}
\affiliation{%
INESC Microsistemas e Nanotecnologia \\
Rua Alves Redol, 9 \\
1000-029 Lisboa, Portugal\\
}

\author{B. Hillebrands}
\affiliation{%
Fachbereich Physik and Forschungsschwerpunkt MINAS\\
Technische Universit{\"a}t Kaiserslautern\\
Erwin-Schr{\"o}dinger-Str.\ 56, 67663 Kaiserslautern, Germany\\
}

\date{\today}

\begin{abstract}

The influence of second-order magneto-optic effects on Kerr effect
magnetometry of epitaxial exchange coupled \femn/\nife-bilayers is
investigated. A procedure for separation of the first- and
second-order contributions is presented. The full angular
dependence of both contributions during the magnetization reversal
is extracted from the experimental data and presented using gray
scaled magnetization reversal diagrams. The theoretical
description of the investigated system is based on an extended
Stoner-Wohlfarth model, which includes an induced unidirectional
and fourfold anisotropy in the ferromagnet, caused by the coupling
to the antiferromagnet. The agreement between the experimental
data and the theoretical model for both the first- and
second-order contributions are good, although a coherent reversal
of the magnetization is assumed in the model.
\end{abstract}

%\pacs{75...}

\maketitle

%------ Begin Text -------%

\section{Introduction}
Since its discovery in 1877 by J. Kerr \cite{kerr1877} the
magneto-optic Kerr effect (MOKE) has evolved into a very powerful
tool for characterization of magnetic materials. Due to its high
sensitivity MOKE magnetometry is widely used for thin film and
multilayer analysis. The high lateral resolution of modern MOKE
magnetometry enables the study of individual magnetic
nanostructures
\cite{cowburn2000,cowburn2002,allwood2002,allwood2003}. Recent
developments using stroboscopic magneto-optic techniques achieved
high time resolution
\cite{freeman1992,crawford1999,bauer2000,silva2002}, thus enabling
the study of the magnetization dynamics on a picosecond-time
scale. Using second harmonic generation in MOKE measurements
results in a high sensitivity to the magnetization at the
interfaces between different materials
\cite{pan1989,huebner1989,reif1991,reif1993,bennemann1998}.

The origin of magneto-optic effects is the spin-orbit interaction.
In many cases it is sufficient to treat the magneto-optic response
in first order, i.e. take into account only contributions linearly
proportional to the magnetization. However as first shown by
Osgood {\sl et al.} \cite{osgood1998} second-order magneto-optic
effects can be important in thin films with in-plane anisotropy.
In particular for magnetization reversal measurements using MOKE
magnetometry the second-order contributions can lead to asymmetric
hysteresis loops
\cite{zhong1990,bland1990,osgood1995,postava1997,osgood1998},
which are not observed using other magnetometry methods. On the
other hand in exchange bias systems, which consist of a
ferromagnet exchange coupled to an antiferromagnet, asymmetric
hysteresis loops have been reported independently of the
magnetometry method
\cite{ambrose1998,nogues1999,leighton2000,fitzsimmons2000,gierlings2002,krivorotov2002,mccord2003}.
Therefore special care is necessary when investigating exchange
bias systems using magneto-optical Kerr effect magnetometry in
order to distinguish between the effects caused by second-order
magneto-optics and those caused by the broken symmetry due to the
exchange bias effect.

In this article we use the epitaxial \femn/\nife\ exchange bias
model system to show how second-order magneto-optic effects affect
the magnetization reversal observed in MOKE magnetometry. By
utilizing a simple procedure described in this article both the
first- and second-order effects can easily be separated. The
experimental data is summarized and compared with an extended
Stoner-Wohlfarth model using magnetization reversal diagrams. Our
approach builds upon a method to extract the linear and the
quadratic Kerr contributions from Kerr effect measurements, which
has recently been proposed by Mattheis et al., and in which a
magnetic field of constant field strength is rotated about the
axis normal to the sample surface, ("ROTMOKE" method,
\cite{mattheis1999,mattheis199b}). In contrast to this method,
which is reminiscent to a torque measurement, the method proposed
in the current article is based on the analysis of the
magnetization reversal of the sample under investigation.

\section{Experiment}
The samples were prepared in an UHV system with a base pressure of
$5\times 10^{-11}\,$ mbar. In order to epitaxially grow
\femn/\nife\ bilayers single crystalline MgO(001) substrates were
used, first depositing a buffer layer system consisting of
Fe(0.5\,nm)/Pt(5\,nm)/Cu(100\,nm) described in detail elsewhere
\cite{mewes2001}. The samples consist of a 10\,nm thick \femn\
layer and a 5\,nm thick \nife\ layer covered by 2\,nm Cu in order
to ensure symmetric interfaces and by 1.5\,nm Cr to prevent
oxidation. The different materials were evaporated using either an
e-beam evaporator (Fe, Pt, \nife, Cr) or Knudsen cells (Cu, Mn),
with typical evaporation rates ranging from 0.01\,nm/s to
0.1\,nm/s. The layer composition and crystallographic structure
was characterized using a combined low energy electron diffraction
(LEED) and Auger system. Further structural investigation was
performed using reflecting high energy diffraction (RHEED) and
{\em in-situ} scanning tunneling microscopy (STM). The samples
were heated after deposition in UHV slightly above the bulk
N\'{e}el-temperature of \femn (500\, K), while a magnetic field of
500\, Oe was applied along the in-plane [100]-direction of \nife\
during cool down.

\section{Results and discussion}
\begin{figure}[tb!]
\includegraphics[angle=-90,width=\columnwidth,clip]{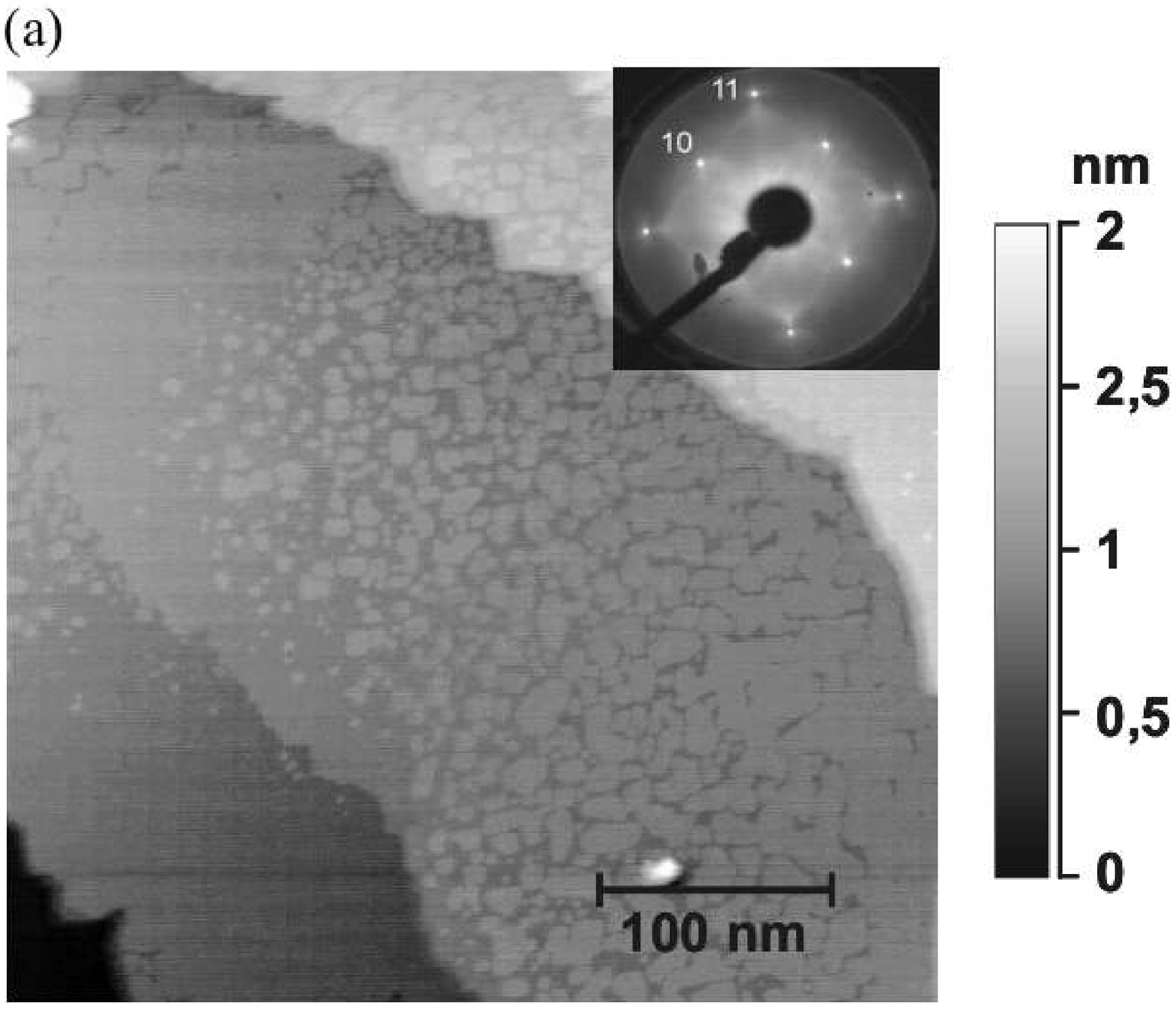}
\includegraphics[angle=-90,width=\columnwidth,clip]{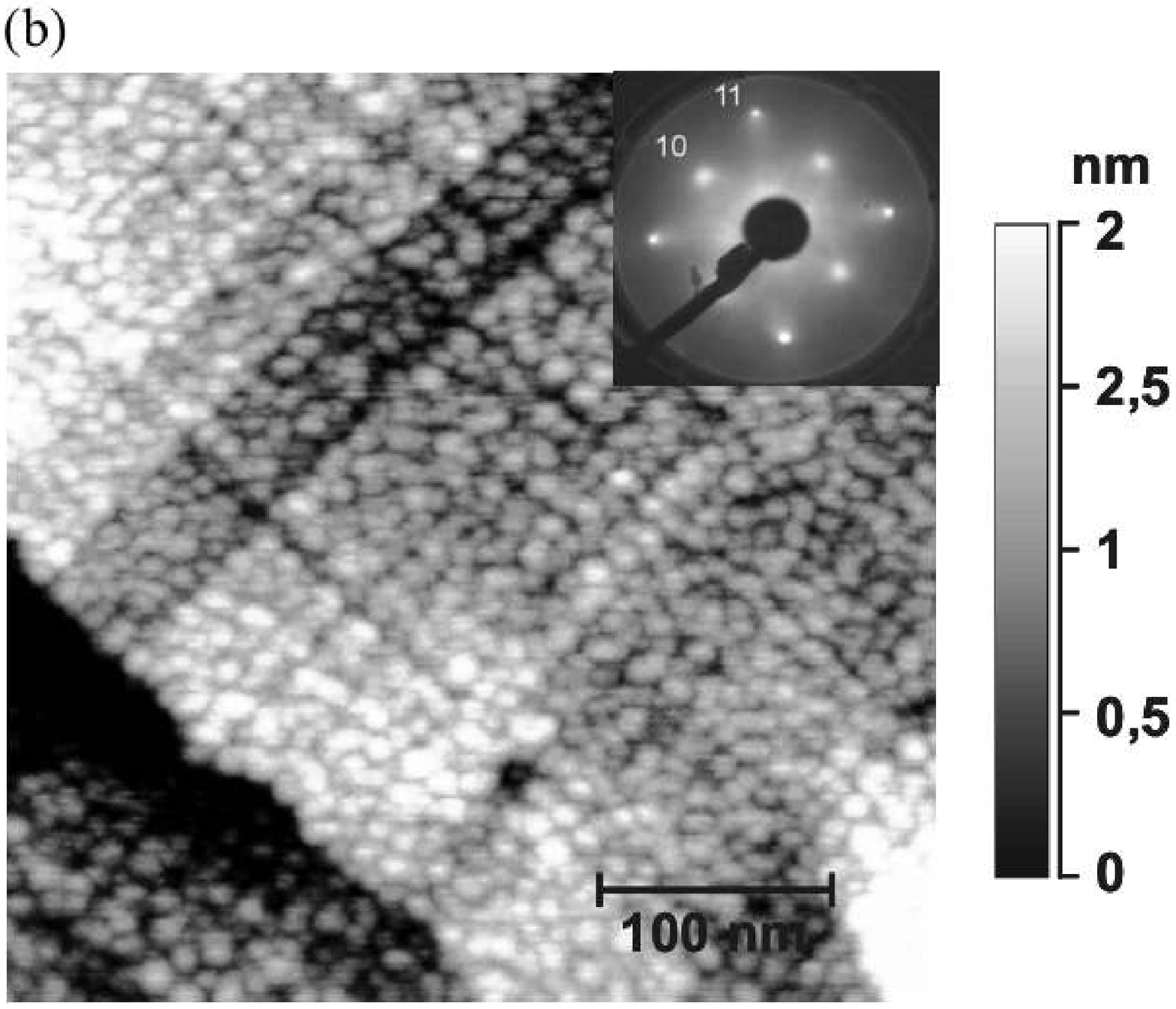}
\caption{(a) STM image of a 10\,nm thick (001)-oriented
\femn-layer grown on MgO(001)/Fe/Pt/Cu, the scan area is $0.4\,
\mu$m$\times 0.4\, \mu$m, with a full height scale of 2\,nm. The
inset shows the LEED pattern of the same surface at a primary
energy of 109\, eV. (b) STM image of a 5\,nm thick (001)-oriented
\nife-layer grown on top of a 10\,nm thick \femn-layer, the scan
area is $0.4\, \mu$m$\times 0.4\, \mu$m, with a full height scale
of 2\,nm. The inset shows the LEED pattern of the same surface at
a primary energy of 128\, eV. } \label{stm}
\end{figure}

A \femn\ layer deposited on top of the Cu(001) buffer layer by
co-evaporation of Fe by e-beam evaporation and Mn from a Knudsen
cell also grows in (001) orientation, with
$[100]_{FeMn}||[100]_{Cu}$. The surface morphology consists of
rather large terraces with small monoatomic islands on top. These
small islands have a large size distribution, as can be seen in
the STM image in Fig. \ref{stm} (a). \nife\ deposited on
\femn(001) also grows in (001) orientation but shows a broadening
of the LEED spots due to formation of small islands with an
average size of 10\,nm, while the larger terraces of the
underlying \femn\ are still visible, as can be seen in Fig.
\ref{stm} (b).

\begin{figure}[tb!]
\includegraphics[clip,angle=-90,width=\columnwidth]{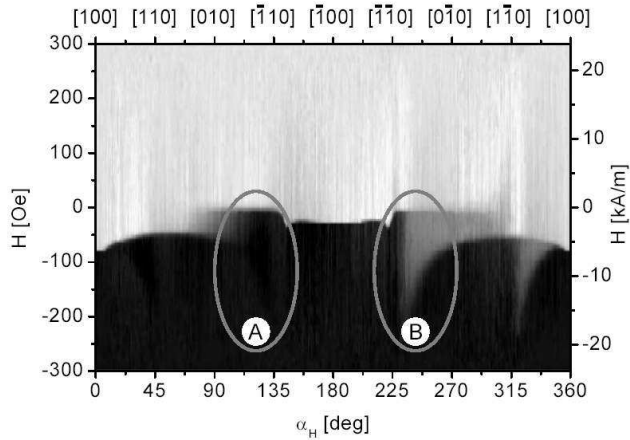}
\caption{Magnetization reversal diagram for the branch of the
hysteresis curve with decreasing external magnetic field of an
epitaxial \nife/\femn\ sample, as measured using Kerr effect
magnetometry. The grayscale is proportional to the Kerr-rotation.
The regions where the asymmetry discussed in the text is most
obvious are marked 'A` and 'B`.} \label{kerroriginal}
\end{figure}

The magnetic properties of a \femn(10\, nm)/\nife(5\, nm) bilayer
are measured using Kerr effect magnetometry. The magnetic field is
applied collinear to the plane of the incident s-polarized light.
The angle $\alpha_H$ of the in-plane [100]-direction of the \nife\
layer relative to the plane of the incident light is varied from 0
to 360\,degree in 1 degree steps by rotating the sample. For all
experimental data obtained from this rotation the decreasing field
branch is shown in Fig. \ref{kerroriginal}, using a magnetization
reversal diagram with a grayscale proportional to the
Kerr-rotation. This kind of data visualization enables the
presentation of the whole angular dependence of the magnetization
reversal in a single diagram and was described in detail elsewhere
\cite{mewes2002}. As can be seen in this figure the magnetization
reversal diagram of the \femn /\nife\ exchange bias system shows
an asymmetry, which is characteristic for quadratic contributions
to the Kerr rotation, as will be shown in the following. This
asymmetry impedes a correct determination of the angular
dependence of the coercive field and the exchange bias field from
the raw data causing those quantities to be asymmetric with
respect to the in-plane angle $\alpha_H$.

\begin{figure}[tb!]
\includegraphics[angle=-90,width=\columnwidth,clip]{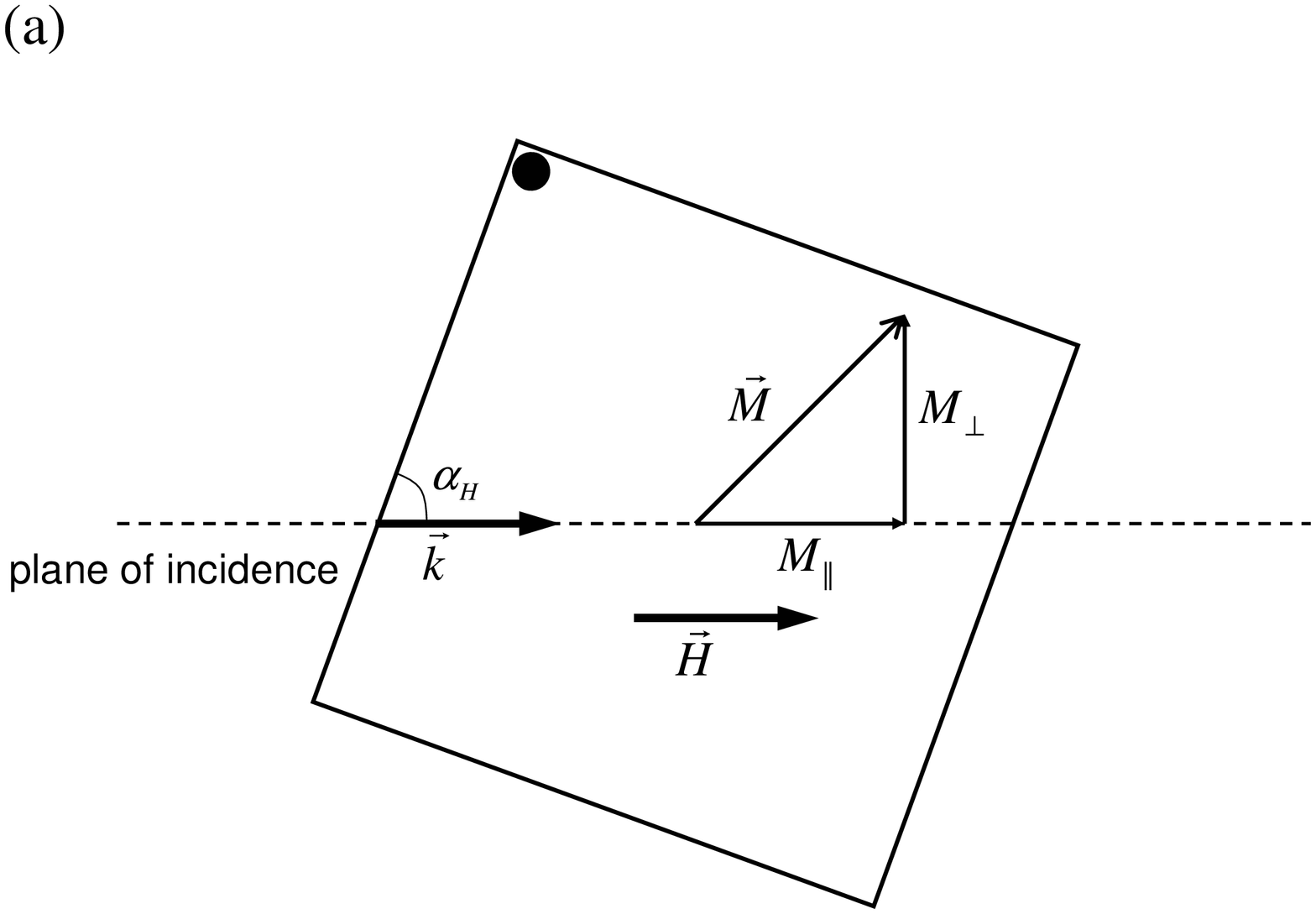}
\includegraphics[angle=-90,width=\columnwidth,clip]{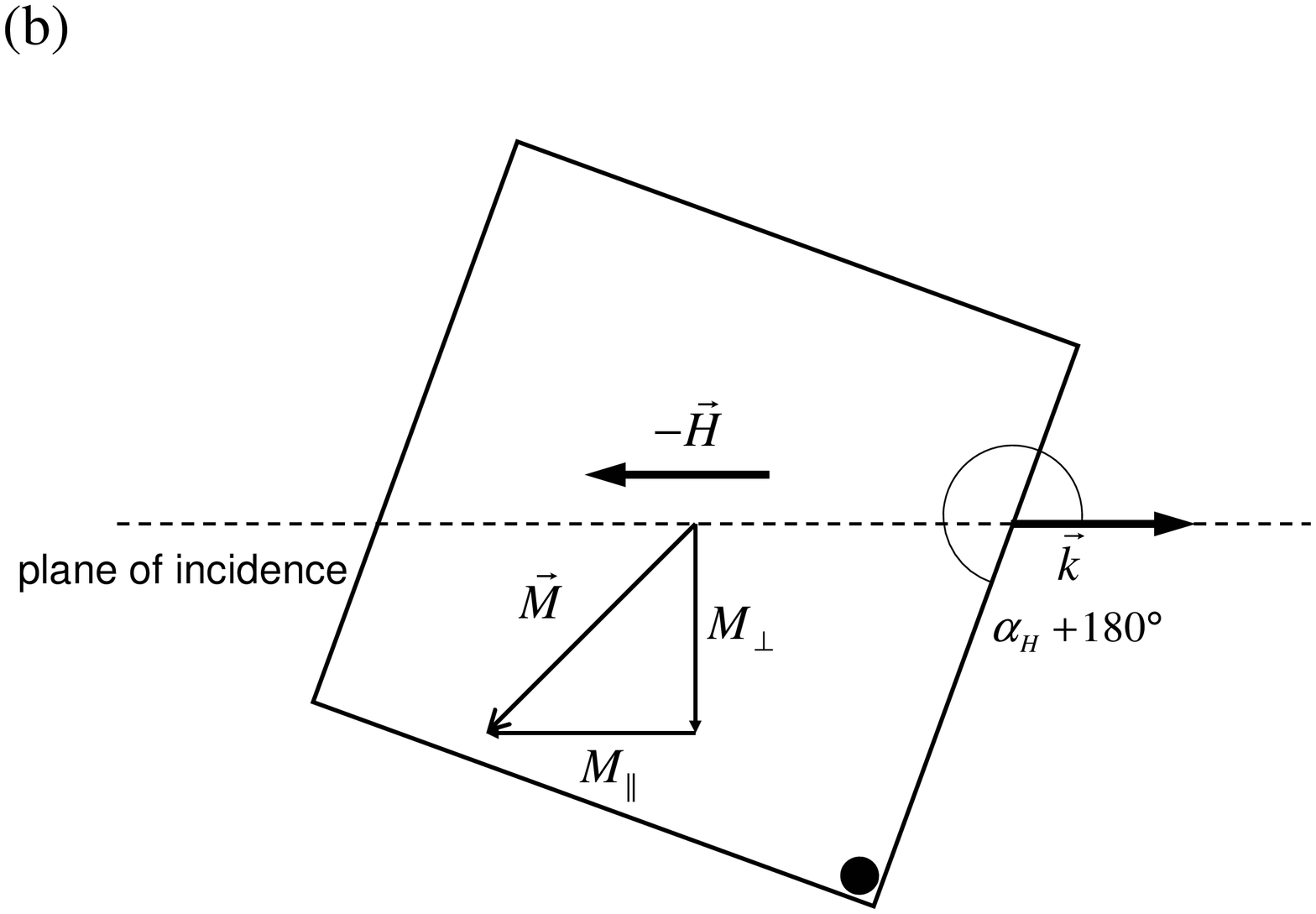}
\caption{Geometry used to separate the different contributions to
the Kerr rotation $\theta_{Kerr,s}$. (a) Situation for an angle
$\alpha_H$ of the [100]-direction with respect to the plane of the
incident light (characterized by the wavevector $\vec{k}$). (b)
Equivalent situation with rotation of the sample by 180\,deg and
reversed direction of the applied magnetic field. The filled
circle marks the orientation of the sample.} \label{geometry}
\end{figure}

\begin{figure}[tb!]
\includegraphics[angle=-90,width=\columnwidth,clip]{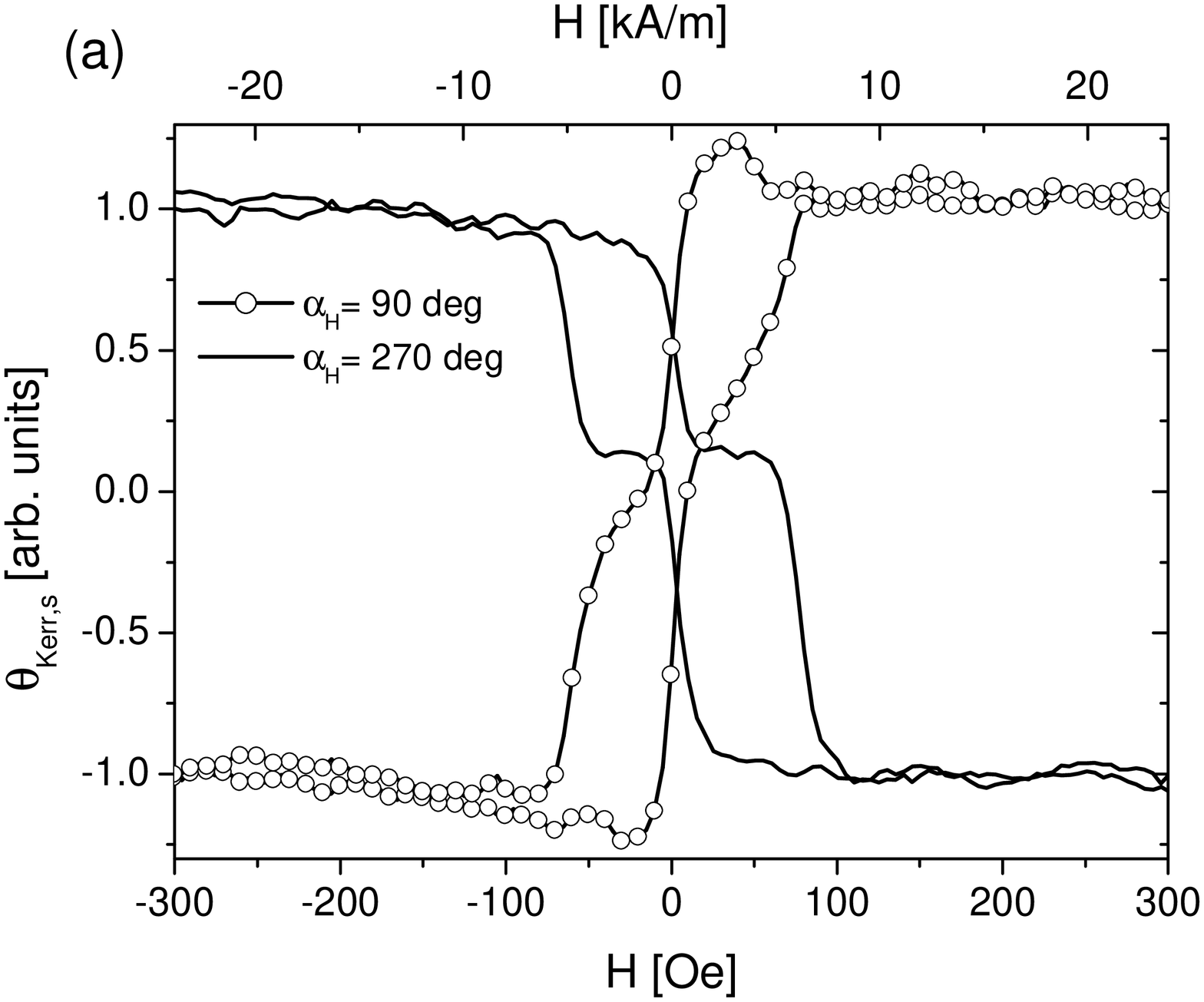}\\
\includegraphics[angle=-90,width=\columnwidth,clip]{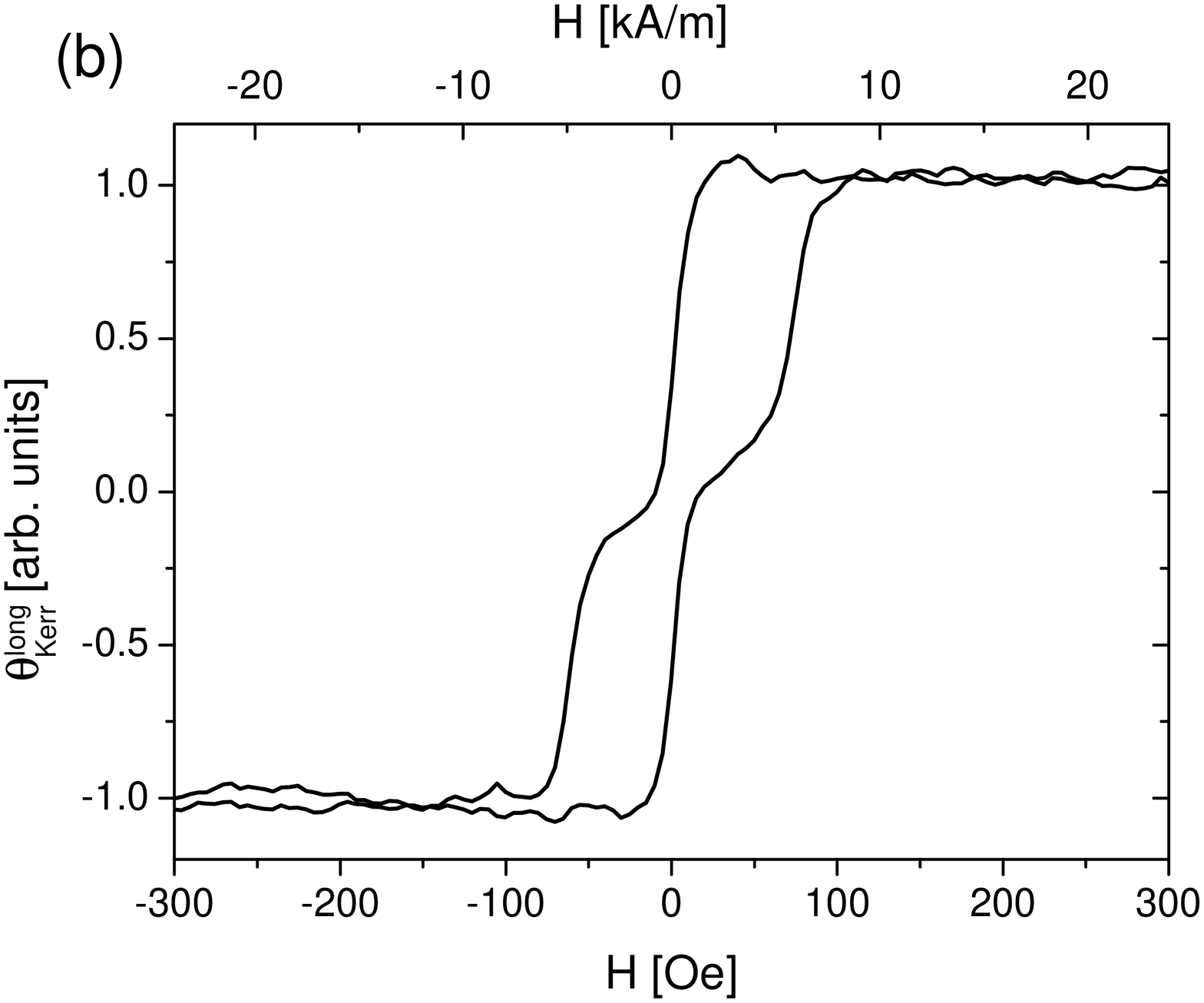}
\includegraphics[angle=-90,width=\columnwidth,clip]{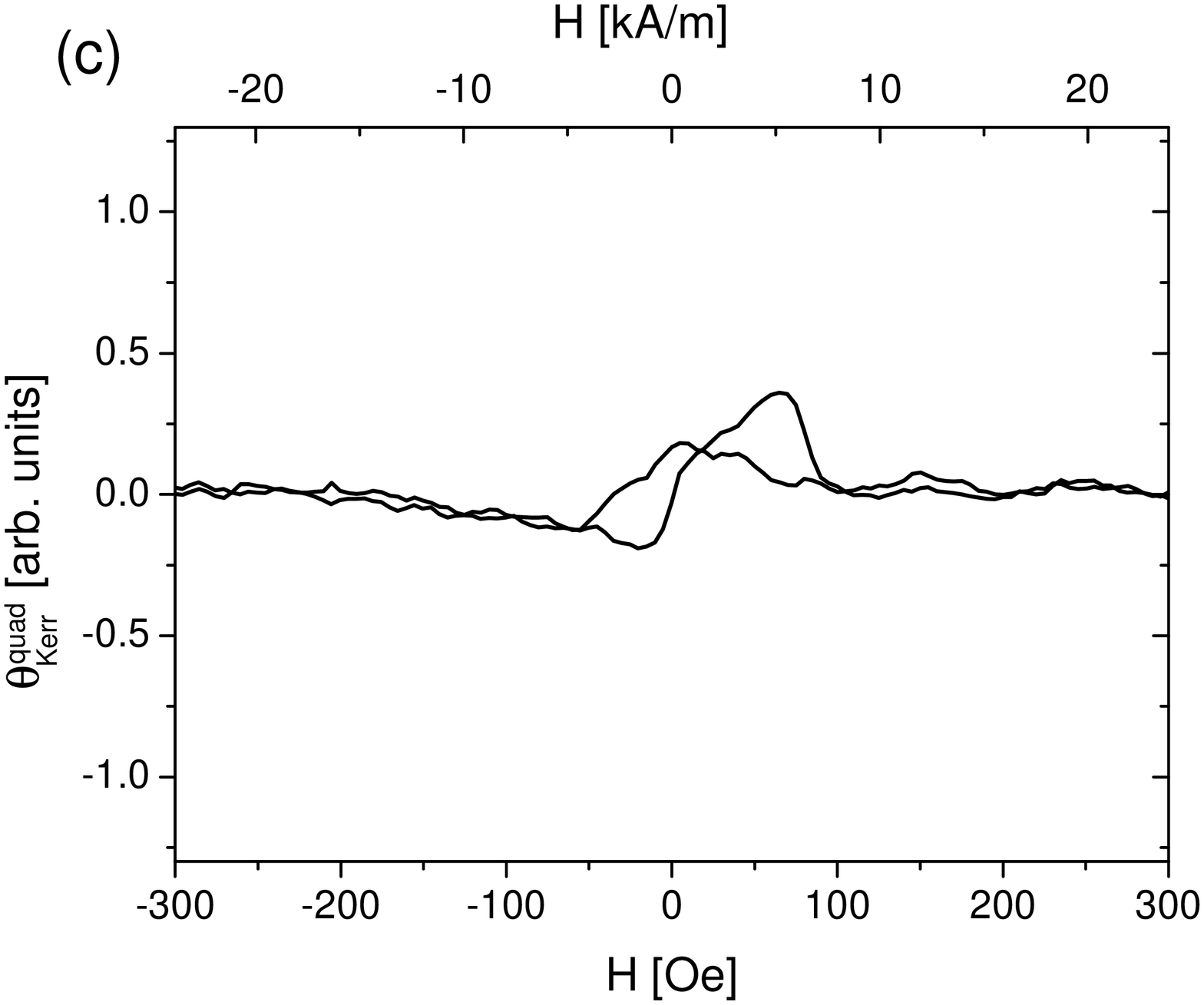}
\caption{(a) Magnetization reversal for two equivalent angles
$\alpha_H=90\,$deg (open symbols) and $\alpha_H=270\,$deg (line).
Note that the sign of the magnetic field for the magnetization
reversal at $\alpha_H=270\,$deg was reversed, so that the reversal
is equivalent to that at $\alpha_H=90\,$deg. In (b) the linear
longitudinal contribution $\theta_{Kerr}^{long}$ to the Kerr
rotation of the magnetization reversal in (a) is shown, while in
(c) the second-order contribution $\theta_{Kerr}^{quad}$ is
shown.} \label{reversal}
\end{figure}

The Kerr rotation $\theta_{Kerr,s}$ in longitudinal geometry with
s-polarized light and the sample magnetized in the plane of the
sample surface, can be written as follows
\cite{postava1997,mattheis1999,mattheis199b,lopusnik2001,postava2002}:
\begin{equation}\label{kerr1}
\theta_{Kerr,s}=\vartheta_{Kerr}^{long}M_{\|}+\vartheta_{Kerr}^{quad}M_{\|}M_{\perp},
\end{equation}
where $M_{\|}$ and $M_{\perp}$ are the in-plane magnetization
components parallel and perpendicular to the plane of incidence of
the light. $\vartheta_{Kerr}^{long}$ and $\vartheta_{Kerr}^{quad}$
are the longitudinal and quadratic proportionality factors of the
Kerr rotation. The second order term proportional to the product
of the longitudinal and transverse component is the reflection
analogy of the Voigt effect
\cite{lissberger1971,lissberger1971b,carey1974,postava1997} and
gives rise to the asymmetry observed in Fig. \ref{kerroriginal}.
The two contributions to the Kerr rotation can be separated by
making use of the symmetry of the problem as follows. As
illustrated in Fig. \ref{geometry}, if the in-plane angle
$\alpha_H$ of the sample with respect to the plane of the incident
light is changed by 180\,deg and the sign of the magnetic field
$\vec{H}$ is reversed the same magnetization reversal process
should be observed.
However by doing so the first term in equation (\ref{kerr1})
proportional to $M_{\|}$ changes sign while the second term
proportional to $M_{\|}M_{\perp}$ will have the same sign for both
sample orientations. This leads to apparently different
magnetization reversal curves observed in Kerr effect
magnetometry, an example of which is shown in Fig. \ref{reversal}
(a).

By calculating the difference of the magnetization reversal for
$\alpha_H$ and $\alpha_H+180\,$deg the Kerr rotation
$\theta_{Kerr}^{long}$ caused by the longitudinal component of the
magnetization $M_{\|}$ can be reconstructed:
\begin{eqnarray}
\theta_{Kerr}^{long}&:=&[\theta_{Kerr}(\alpha_H)-\theta_{Kerr}(\alpha_H+180^\circ)]/2\\
&=&\vartheta_{Kerr}^{long}M_{\|}.\nonumber
\end{eqnarray}
This is shown in Fig. \ref{reversal} (b) for the magnetization
reversals shown in part (a) of the same figure. On the other hand
the quadratic contribution $\theta_{Kerr}^{quad}$ to the Kerr
rotation can be obtained by calculating the average of the Kerr
rotation at $\alpha_H$ and $\alpha_H + 180\,$deg:
\begin{eqnarray}
\theta_{Kerr}^{quad}&:=&[\theta_{Kerr}(\alpha_H)+\theta_{Kerr}(\alpha_H+180^\circ)]/2\\
&=& \vartheta_{Kerr}^{quad}M_{\|}M_{\perp},\nonumber
\end{eqnarray}
as shown in Fig. \ref{reversal} (c).

\begin{figure}[tb!]
\includegraphics[angle=-90,width=\columnwidth,clip]{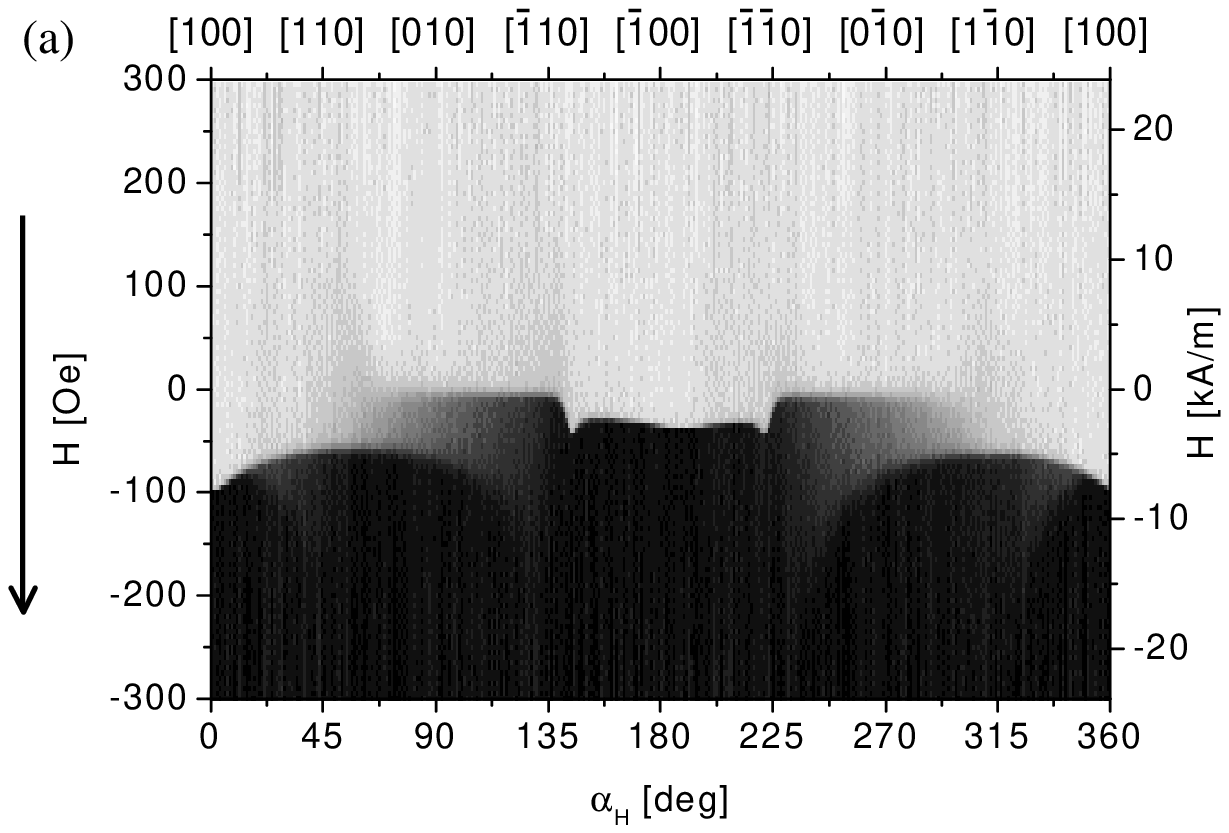}
\includegraphics[angle=-90,width=\columnwidth,clip]{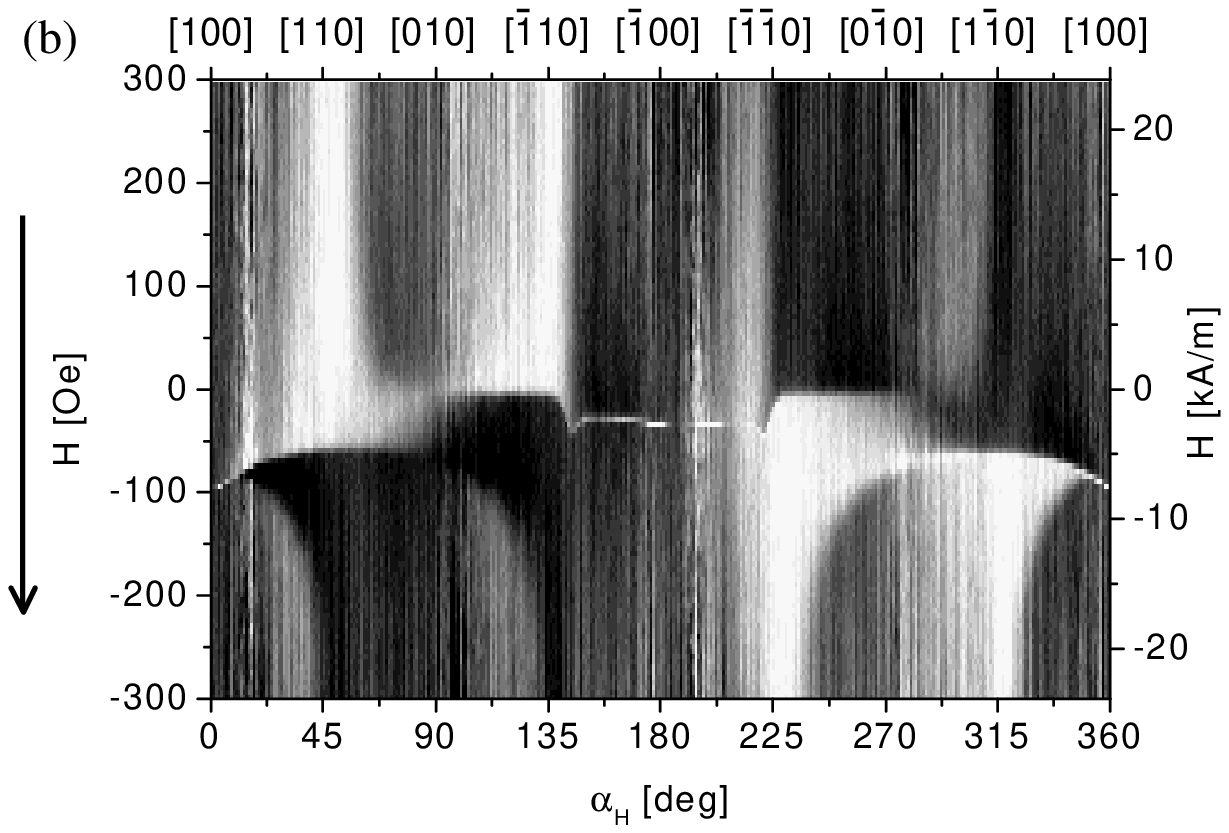}
\caption{a) Measured magnetization reversal diagram for
$\theta_{Kerr}^{long}$ caused by the longitudinal component of the
magnetization $M_{\|}$. In b) the corresponding reversal diagram
of the second-order contribution $\theta_{Kerr}^{quad}$ caused by
$M_{\|}M_{\perp}$ is shown. In both graphs the grayscales are
chosen differently in order to fit the respective data range.}
\label{revdiag}
\end{figure}

\begin{figure}[tb!]
\includegraphics[angle=-90,width=\columnwidth,clip]{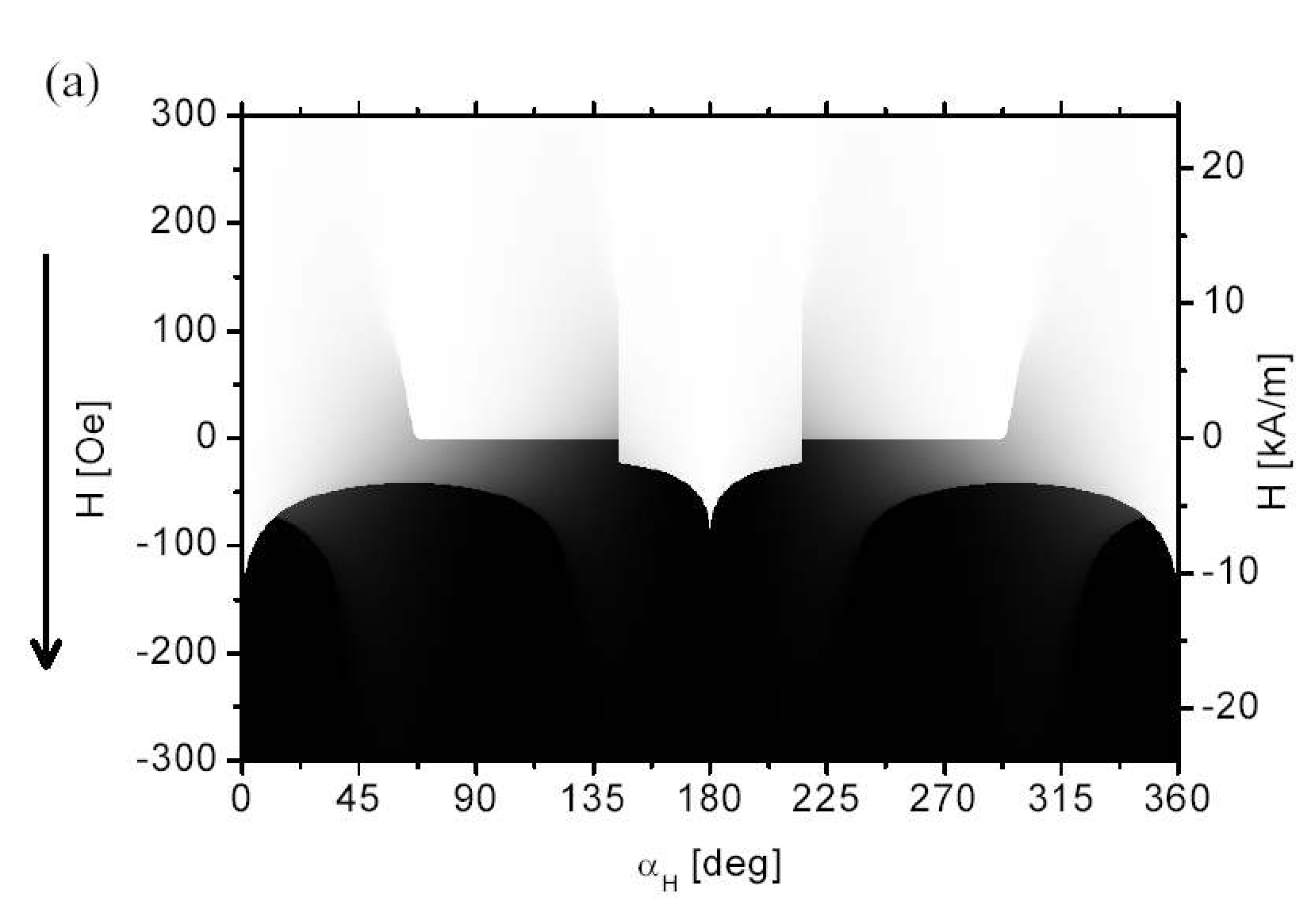}
\includegraphics[angle=-90,width=\columnwidth,clip]{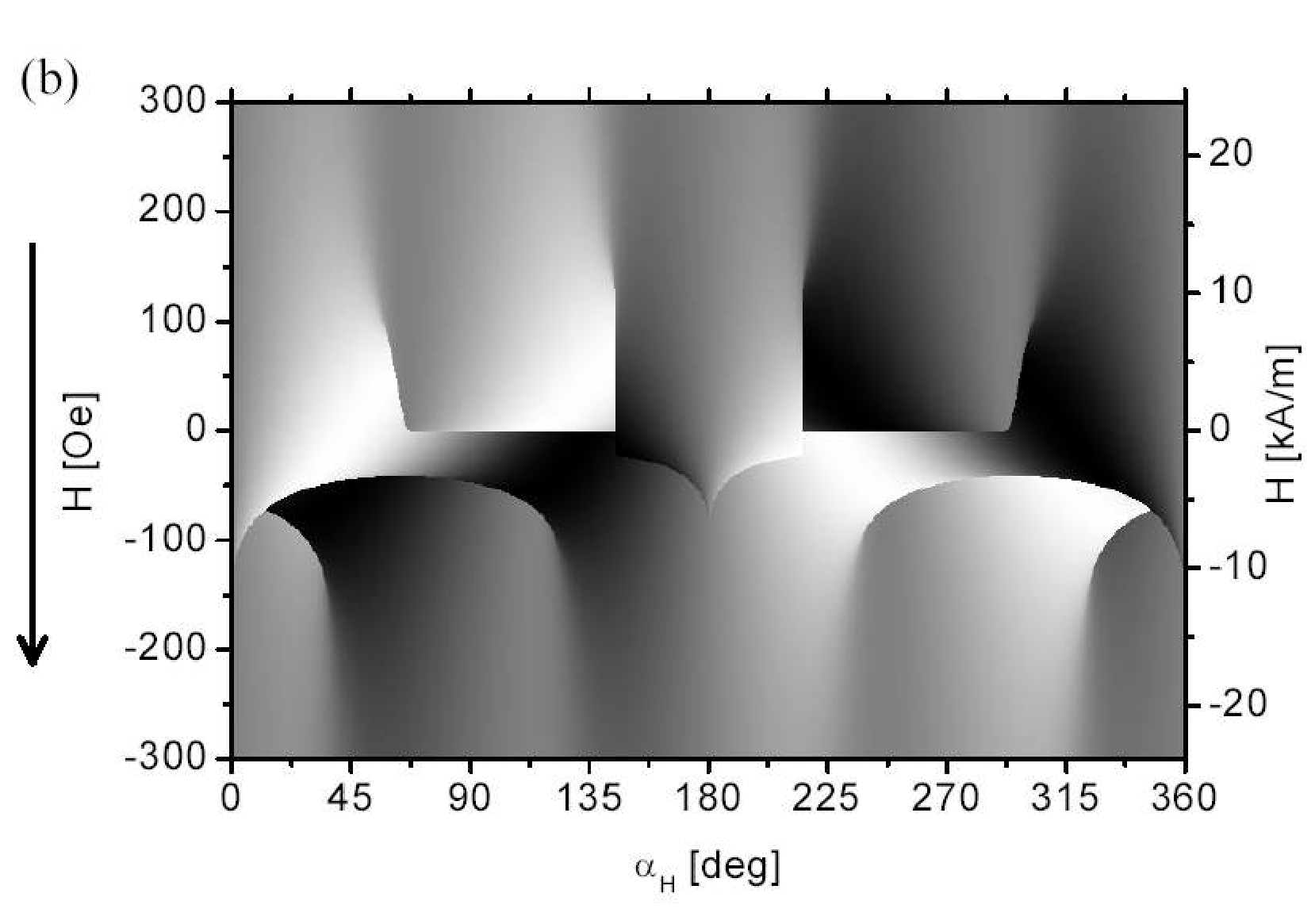}
\caption{Magnetization reversal diagrams of the decreasing field
branch, as predicted by the extended Stoner-Wohlfarth model using
equation \ref{freeenergy}, with $K_1=2.7\,$erg/$\text{cm}^3$ and
$K_4=4.9\,$erg/$\text{cm}^3$. In a) the longitudinal component
$M_{\|}$ is shown, while in b) the grayscale is proportional to
the product $M_{\|}M_{\perp}$.} \label{theoreversal}
\end{figure}

By carrying out the same kind of analysis for all angles
$\alpha_H$ the magnetization reversal diagram for
$\theta_{Kerr}^{long}$, i.e. for the longitudinal component of the
magnetization, can be reconstructed, as is shown in Fig.
\ref{revdiag} (a). Consequently in this figure the asymmetry that
was observed in Fig. \ref{kerroriginal} is no longer present. Note
however, that the symmetry breaking effect of the exchange bias
effect is still visible in this diagram. A similar diagram can be
constructed for the quadratic contribution $\theta_{Kerr}^{quad}$
to the Kerr rotation, as shown in Fig. \ref{revdiag} (b). As this
diagram contains information about the product $M_{\|}M_{\perp}$
it reflects the corresponding symmetry (see also Fig.
\ref{theoreversal} (b) discussed later).

\begin{figure}[tb!]
\includegraphics[angle=-90,width=\columnwidth,clip]{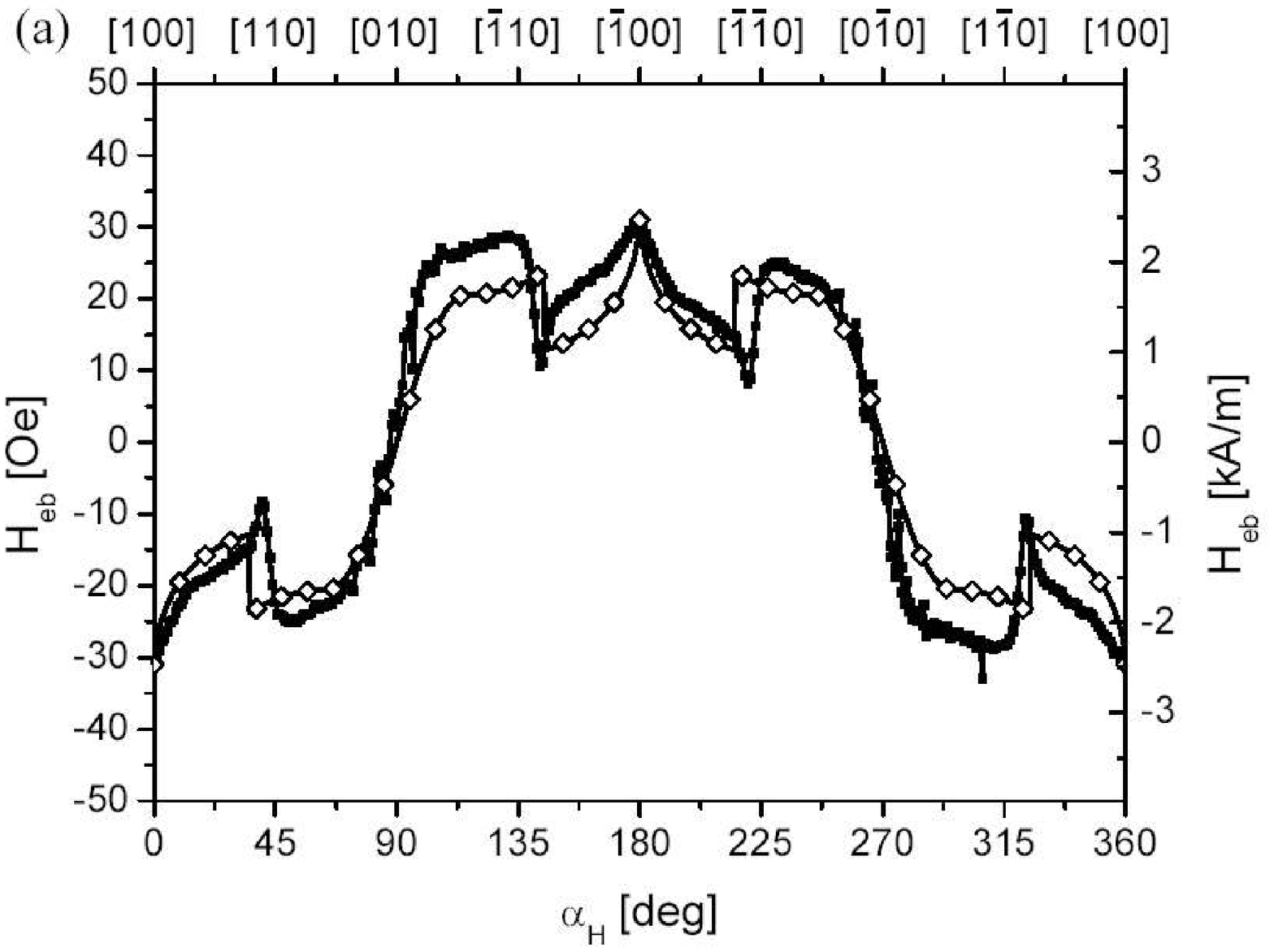}
\includegraphics[angle=-90,width=\columnwidth,clip]{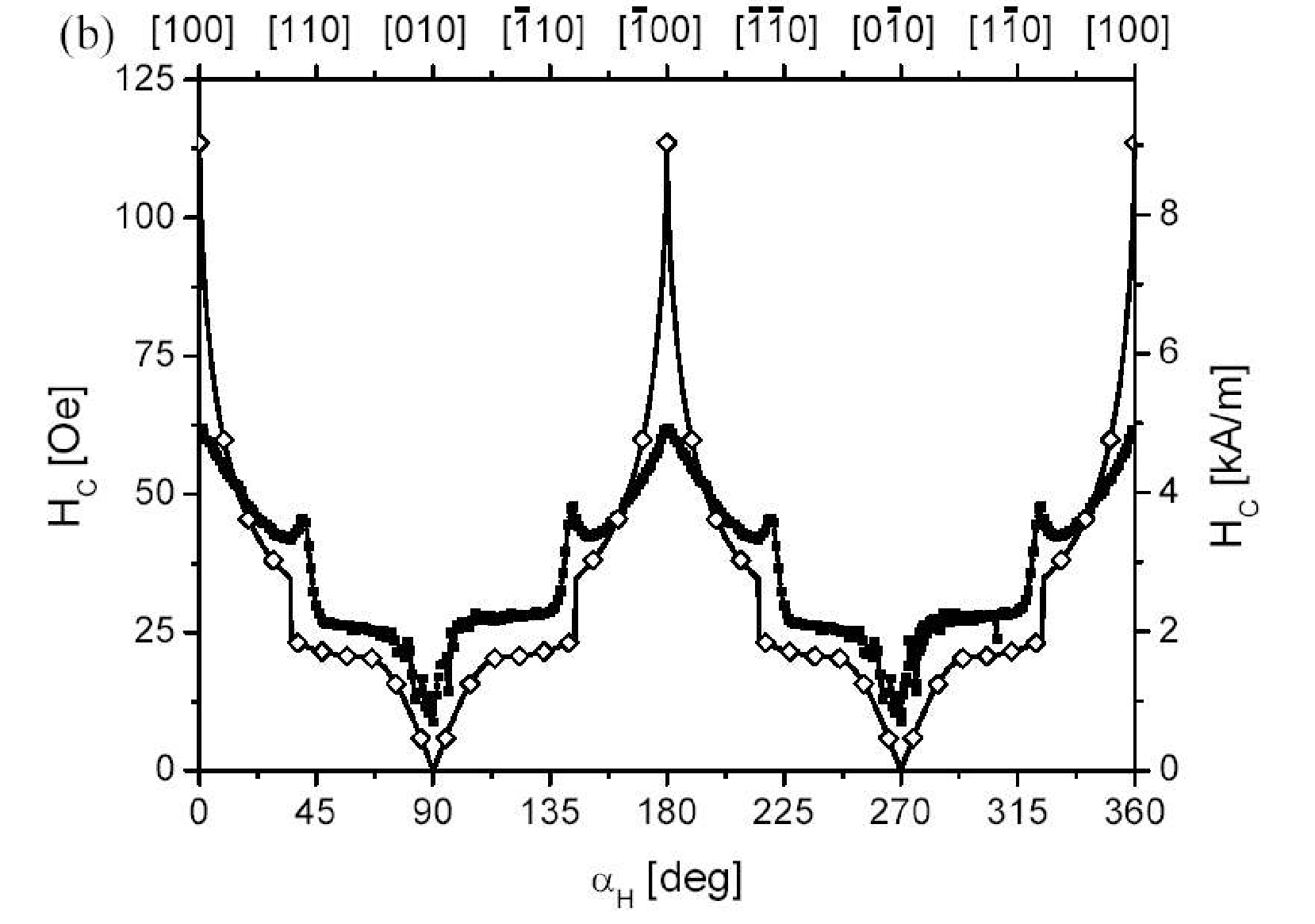}
\caption{Angular dependence of (a) the exchange bias field and (b)
the coercivity. The experimental data is shown as solid symbols,
while the fit using equation \ref{freeenergy} is shown as open
symbols.} \label{hebhc}
\end{figure}

The reversal data of the longitudinal component of the
magnetization in Fig. \ref{revdiag} (a) is used to derive the
angular dependence of the exchange bias field $H_{eb}$ (see Fig.
\ref{hebhc} (a)) and the coercive field $H_C$ (see Fig.
\ref{hebhc} (b)) of the \femn /\nife\ double layer system. These
angular dependencies are then fitted assuming a coherent rotation
of the magnetization and using the perfect-delay convention
\cite{nieber1991} within the framework of an extended
Stoner-Wohlfarth model \cite{mewes2002,mewes2003}.
The experimental data can be reasonably well described by
including a unidirectional anisotropy $K_1$ and a fourfold
anisotropy $K_4$ contribution to Gibb's free energy $g$ of the
system, which in turn can be written as:
\begin{eqnarray}\label{freeenergy}
 g=&-&K_1 \cos(\alpha_M)+K_4
\sin^2(\alpha_M)\cos^2(\alpha_M)\\
&-&HM_S\cos(\alpha_M-\alpha_H).\nonumber
\end{eqnarray}
A fit of the experimental data shown in Fig. \ref{hebhc} using the
Gibb's free energy given by equation \ref{freeenergy} results in a
unidirectional anisotropy $K_1=(2.7\pm 0.1)\,$erg/$\text{cm}^3$
and a fourfold anisotropy $K_4=(4.9\pm 0.2)\,$erg/$\text{cm}^3$.
Note that the appearance of an induced fourfold anisotropy in
addition to the unidirectional anisotropy in epitaxial
\femn/\nife-bilayer systems has recently been shown theoretically
using a vector spin model \cite{mewes2003b}. The resulting angular
dependence of the exchange bias field $H_{eb}$ and the coercive
field $H_C$ predicted by the extended Stoner-Wohlfarth model is
also shown in Fig. \ref{hebhc}.

In order to complete the picture of the magnetization reversal
that results from these anisotropies within the extended
Stoner-Wohlfarth model in Fig. \ref{theoreversal} the reversal
diagrams are given for both $M_{\|}$ and $M_{\|}M_{\perp}$. These
two diagrams correspond to the expected linear and second-order
contribution to the magneto-optic Kerr effect respectively and can
therefore be directly compared with the experimental results in
Fig. \ref{revdiag}.
Given the simplification of a coherent magnetization reversal
process assumed in the extended Stoner-Wohlfarth model and the
small number of fitting parameters the agreement between the model
and the experimental results is surprisingly good. However one
notices differences between the model calculations and the
experimental results especially along the axis parallel to the
easy direction of the unidirectional anisotropy, i.e. around
0\,deg and 180\,deg. Similar deviations have been observed in
epitaxial NiFe/FeMn bilayers \cite{mewes2002} (i.e. in a system
with reversed layer sequence) and may be related to thermal
activation \cite{mewes2003} or to domain formation and
propagation, which are not taken into account in the
Stoner-Wohlfarth model.

\section{Summary}
In summary we have shown that second-order magneto-optic effects
are present in exchange coupled epitaxial \femn/\nife-bilayers. By
using the method described in this article it is possible to
separate the first- and second-order contributions. Thereby the
asymmetry related to magneto-optics can also be separated from the
one associated with the exchange bias effect. The experimental
data can thus be analyzed within an extended Stoner-Wohlfarth
model, which describes well the overall angular dependence of the
magnetization reversal. The observed differences between the
experimental data and the Stoner-Wohlfarth model may be caused by
thermal activation or domain formation and propagation.
\begin{acknowledgments}
We would like to thank R. Lopusn\'{\i}k for stimulating and
helpful discussions.
\end{acknowledgments}
{}
\newpage
\end{document}